\documentclass[aps,prl,reprint,showpacs,longbibliography]{revtex4-1}
\usepackage{amsmath,amssymb,graphicx,units,ulem}
\usepackage[usenames,dvipsnames]{color}

\newcommand{\beginsupplement}{%
        \setcounter{table}{0}
        \renewcommand{\thetable}{S\arabic{table}}%
        \setcounter{figure}{0}
        \renewcommand{\thefigure}{S\arabic{figure}}%
}

\begin{document}
\title{Many-body localization and thermalization in disordered Hubbard chains}
\author{Rubem Mondaini}
\author{Marcos Rigol}
\affiliation{Department of Physics, The Pennsylvania State University, University Park, Pennsylvania 16802, USA}

\begin{abstract}
We study the many-body localization transition in one-dimensional Hubbard chains using exact diagonalization and quantum chaos indicators. We also study dynamics in the delocalized (ergodic) and localized phases and discuss thermalization and eigenstate thermalization, or the lack thereof, in such systems. Consistently within the indicators and observables studied, we find that ergodicity is very robust against disorder, namely, even in the presence of weak Hubbard interactions the disorder strength needed for the system to localize is large. We show that this robustness might be hidden by finite size effects in experiments with ultracold fermions.
\end{abstract}

\maketitle

\paragraph{Introduction.}
Over the years, substantial attention has been devoted to understanding the dynamical properties of disordered systems. Interest on this topic goes back to a seminal paper by Anderson in 1958, who showed that sufficiently strong quenched disorder can produce localization of noninteracting particles, precluding transport in the thermodynamic limit~\cite{Anderson1958}. Destructive interference is at the heart of this phenomenon. It is more prominent in lower dimensions, and, as a result, any nonzero disorder strength leads to localization in one and two dimensions~\cite{Abrahams1979}. A fundamental aspect of Anderson localization is that it occurs not only in the ground state but also in (highly) excited states.

Because of the possibility of localization occurring in interacting systems, a phenomenon termed many-body localization (MBL), disordered systems in the presence of interactions have received a lot of attention in recent years. Early perturbative arguments \cite{fleishman80,altshuler_gefen_97,gornyi_mirlin_05,basko06} and numerical simulations in the presence of strong interactions \cite{oganesyan_huse_07,znidaric08,pal10,khatami_rigol_12,bardarson_pollmann_12} have triggered much research on this topic \cite{nandkishore_huse_review_15,altman_vosk_review_15}. The MBL transition has also started to be explored in experiments with ultracold atoms \cite{kondov_mcgehee_15,Schreiber2015,Bordia2015} and ions \cite{Smith2015}.

The contrast between the properties of many-body eigenstates of interacting systems in the presence and absence of MBL makes apparent how remarkable MBL is. In generic isolated systems, interactions make it possible for the system to act as its own ``effective bath''. If taken out of equilibrium, such systems evolve in time in such a way that observables equilibrate and can be described by traditional ensembles of statistical mechanics (i.e., they thermalize). This is just one of the manifestations of a phenomenon known as eigenstate thermalization \cite{Deutsch1991,Srednicki1994,Rigol2008}, which, in short, means that the expectation value of an observable in an eigenstate of a many-body interacting system is the same as that in thermal equilibrium (with the same mean energy as the eigenstate energy). Eigenstate thermalization has been shown to occur in several many-body quantum systems \cite{Rigol2008,rigol_09a,rigol_09b,santos_rigol_10b,neuenhahn_marquardt_12,khatami_pupillo_13,steinigeweg_herbrych_13,beugeling_moessner_14,kim_14,sorg_vidmar_14}. It is known not to occur only in integrable and MBL systems, i.e., the latter two classes of systems generally do not exhibit thermalization even if they are thermodynamically large \cite{rigol_14,tang_iyer_15}.

As a matter of fact, it was the latter property of MBL systems that was used in the experiments of Ref.~\cite{Schreiber2015} to distinguish between the delocalized (ergodic) regime and the MBL one, for spinful fermions in the presence of a quasi-periodic potential. Motivated by those experiments, in this work we study the MBL transition in Hubbard chains with disorder. We contrast the predictions of quantum chaos indicators for the transition to those from thermalization and eigenstate thermalization. We argue that ergodicity is remarkably robust in these itinerant systems, and show that finite size effects in thermalization indicators might hide this fact in experiments.

\paragraph{Model and the MBL transition.}
To investigate the MBL transition, we use full exact diagonalization and study the Hamiltonian: 
$\hat H = \hat H_0 + \hat H_\text{sb} + \hat H_W$, in which
\begin{eqnarray}
  \hat H_0 = -&t&\sum_{\substack{i=1\\ \sigma=\uparrow,\downarrow}}^{L-1}
  ( \hat c^{\dagger}_{i\sigma} \hat c^{}_{i+1,\sigma}
  + \text{H.c.}) \nonumber \\
  -&t^\prime&\sum_{\substack{i=1\\ \sigma=\uparrow,\downarrow}}^{L-2}
  (\hat c_{i\sigma}^\dagger \hat c_{i+2,\sigma}^{}
  + \text{H.c.}) 
  + U \sum_i^L \hat n_{i\uparrow} \hat n_{i \downarrow},
\label{eq:Hamiltonian}
\end{eqnarray}
is an extended Hubbard model (written in standard notation) in a linear chain of size $L$ (with open boundary conditions), with nearest neighbor hoppings (amplitude $t$), onsite interaction (strength $U$), and next-nearest neighbor hoppings (amplitude $t'$). We have taken $t'\neq0$ so that the model is nonintegrable (quantum chaotic) in the absence of disorder. Additional symmetries, parity, and SU(2), are removed by adding a very weak magnetic field ($h_b$) and chemical potential ($\mu_b$), respectively, at the opposite edges of the chain: $\hat H_\text{sb} =h_b(\hat n_{1,\uparrow}-\hat n_{1,\downarrow}) + \mu_b (\hat n_{L,\uparrow}+\hat n_{L,\downarrow})$ (see Ref.~\cite{supmat} for details). We focus on a uniformly distributed disorder described by $\hat H_W = \sum_{i\sigma}\varepsilon_{i}\hat n_{i\sigma}$, where the local potential $\varepsilon_{i}\in [-{W}/{2},{W}/{2}]$. To contrast the effect of the disorder with the effect of the quasi-periodic potential studied in Refs.~\cite{Schreiber2015}, we also report results for the phase diagram when $\varepsilon_{i}=\frac{\Delta}{2}\cos{(2\pi\beta i+\phi)}$, where $\Delta$ is the potential strength, $\beta=(\sqrt{5}+1)/2$ is the golden ratio, and $\phi$ is an arbitrary phase (as in the Aubry-Andr\'e model \cite{Aubry1980}). Throughout this Rapid Communication, $t=1$ sets the energy scale and $t'=0.5$. We only change $U$ and the disorder strength. The systems studied are at quarter filling, namely,  $N_\uparrow+N_\downarrow=L/2$, with $N_\uparrow\equiv\langle\sum_i^L\hat n_{i\uparrow}\rangle$ and $N_\downarrow\equiv\langle\sum_i^L \hat n_{i\downarrow}\rangle$. We consider two lattice sizes, $L=10$ and 12, where $N_\uparrow=N_\downarrow=L/4$ for $L=12$, and $N_\uparrow=N_\downarrow\pm1$ for $L=10$ \cite{supmat}.

A common quantum chaos indicator used to locate the many-body localization transition in disordered systems is the average ratio between the smallest and the largest adjacent energy gaps, $r_n=\min[\delta^E_n,\delta^E_{n-1}]/\max[\delta^E_n,\delta^E_{n-1}]$, with $\delta^E_n = E_n - E_{n-1}$, and $\{E_n\}$ is the ordered list of energy levels~\cite{oganesyan_huse_07}. Here, in order to reduce finite size effects, we compute the average ratio $\bar r$ over the central half of the spectrum. In the ergodic phase, when the level spacing exhibits a Wigner-Dyson distribution, the average ratio is $r^\text{WD} \approx0.536$, while in the MBL phase, when the level spacing exhibits a Poisson distribution, the average ratio is $r^\text{P}=2\ln2-1\approx0.386$~\cite{Atas2013}.

\begin{figure}[!tb] 
 \includegraphics[width=0.99\columnwidth]{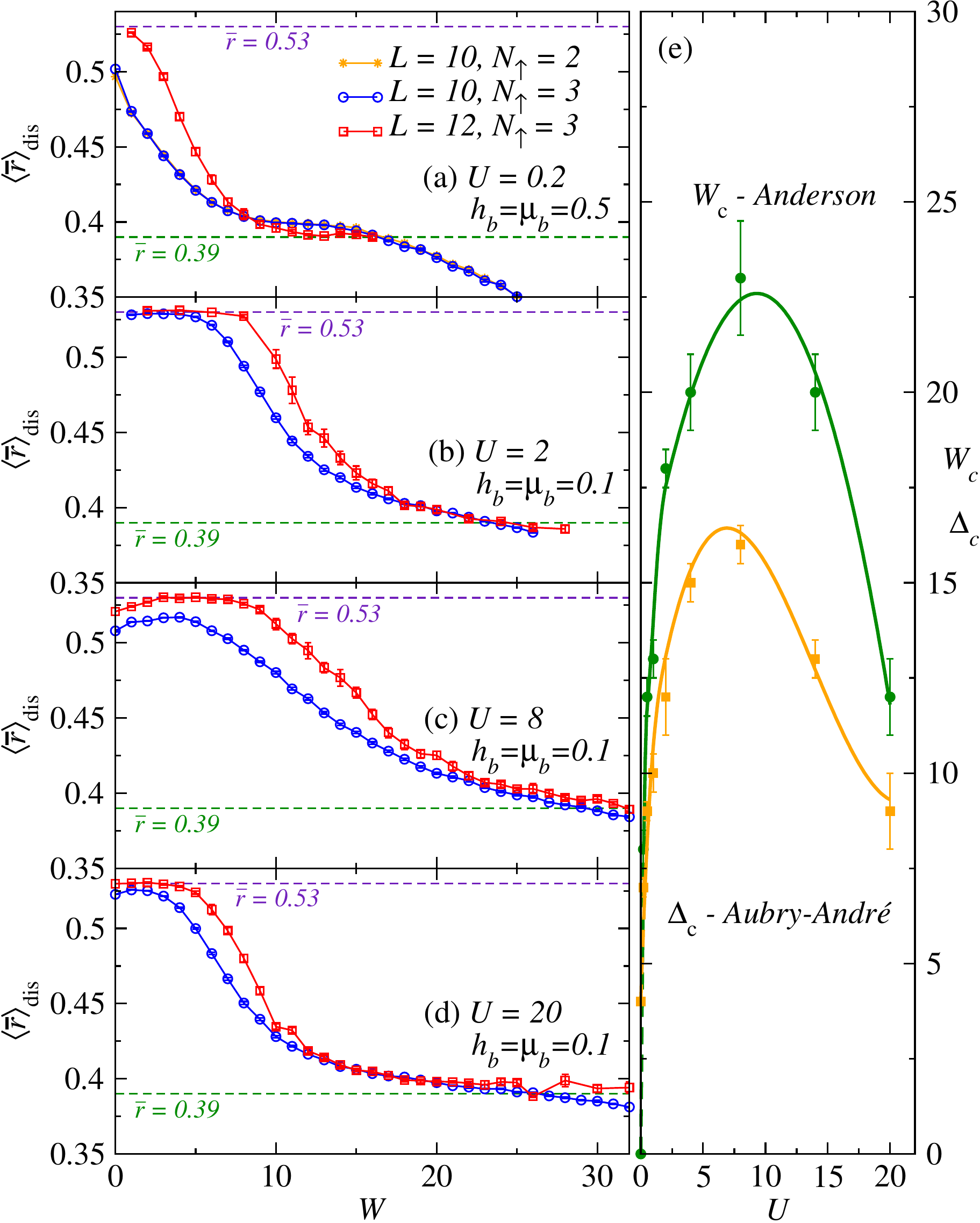}
 \vspace{-0.1cm}
 \caption{(Color online) (a)-(d) Averaged ratio of adjacent energy gaps as a function of the disorder strength for four values of $U$ and two lattice sizes. The average $\bar r$ was computed over the central half of the spectrum. The disordered averaged results $\langle\bar r\rangle_\text{dis}$ for $L=10$ were obtained averaging over $1200$ disorder realizations, and the ones for $L=12$ over 20--200 disorder realizations (error-bars report the standard deviation). In (a), we show results for $N_\uparrow=N_\downarrow\pm1$ when $L=10$. They make apparent that both sectors behave qualitatively (and quantitatively) similarly even for the largest values of $h_b=\mu_b$ used. The crossing (or merging) point between curves for different lattice sizes provides an estimate of the critical disorder, $W_c$, for the ergodic to MBL transition. (e) Estimated $W_c$ and $\Delta_c$ as a function of $U$ (error bars report an interval of confidence based on the closeness of the results for $L=10$ and $L=12$ about $W_c$ and $\Delta_c$).}
 \label{fig:r_w_fixed_tp_v2}
\end{figure}

Figure~\ref{fig:r_w_fixed_tp_v2} shows the disorder average of $\bar r$, $\langle\bar r\rangle_\text{dis}$, as a function of the disorder strength for different values of the on-site repulsion and two system sizes. The value of the disorder strength at which the curves cross or merge, $W_c$, can be taken as an estimate of the critical disorder strength for the ergodic to many-body localization phase transition. Such a crossing or merging point is known to move towards stronger disorder with increasing system size (see, e.g., Refs.~\cite{oganesyan_huse_07,pal10,khatami_rigol_12}); as such, the values reported here should be thought of as lower bounds for the critical disorder. As expected, since interactions promote delocalization, $W_c$ first increases with $U$ [Figs.~\ref{fig:r_w_fixed_tp_v2}(a)--\ref{fig:r_w_fixed_tp_v2}(c)]. It is remarkable that, even for fairly small values of $U$ [$U=0.2$ in Fig.~\ref{fig:r_w_fixed_tp_v2}(a)], the delocalized regime is robust up to values of $W_c\simeq8$, i.e., almost twice the width $B$ of the single particle spectrum, $\epsilon_k=-2t\cos(k)-2t^\prime\cos(2k)$ ($B=4.5$ for our parameters). As the onsite interaction strength becomes of the order of $B$, $W_c$ stops increasing and, as $U$ increases further, $W_c$ starts to decrease [Figs.~\ref{fig:r_w_fixed_tp_v2}(c) and \ref{fig:r_w_fixed_tp_v2}(d)]. This is expected as, in the limit $U\rightarrow\infty$, each sector in the Hubbard model with a particular ordering of the spins (and no double occupancy) maps onto a {\it noninteracting} spinless fermion Hamiltonian with $N_\uparrow+N_\downarrow$ fermions, and the latter localizes for any nonzero disorder strength. Figure~\ref{fig:r_w_fixed_tp_v2}(e) depicts the estimated phase diagram in the presence of disorder for up to $U=20$. In contrast, as also shown in Fig.~\ref{fig:r_w_fixed_tp_v2}(e), MBL in the presence of a quasi-periodic potential (see Ref.~\cite{supmat} for further details) occurs for $\Delta< W$. MBL is also easier to achieve in interacting spinless fermion systems \cite{Lev2015}.

\paragraph{Dynamics and thermalization.}
As mentioned before, one of the defining properties of the MBL phase is its lack of thermalization. In what follows, motivated by the experimental results reported in Ref.~\cite{Schreiber2015}, we study dynamics in the delocalized and MBL regimes. Our initial state is also experimentally motivated. We consider $|\psi_I\rangle=|\uparrow0\downarrow0\uparrow0\downarrow\ldots\rangle$, which is a state that has no double occupancy and can be prepared using optical superlattices. $|\psi_I\rangle$ is a quarter-filling version of the state prepared in Ref.~\cite{Schreiber2015}. The dynamics is then studied under $\hat H = \hat H_0 + \hat H_\text{sb} + \hat H_W$. Our goal is to understand how the results of the dynamics relate to those obtained for $\bar r$. Some of the specific questions we address are the following: Is the MBL transition manifest in the dynamics of experimentally relevant observables? At what time do those observables reach (if they do) stationary values? We are also interested in understanding the role of finite size effects. They have been found to be stronger in indicators related to Hamiltonian eigenstates than in those related to the spectrum \cite{santos_rigol_10a}. To address these questions, we focus on one particular value of the interaction strength, $U=4$. 

We report results for three observables (see Ref.~\cite{supmat} for another one). Two observables, {\it the imbalance} $I=(\langle\hat n^e\rangle-\langle\hat n^o\rangle)/(\langle\hat n^e\rangle+\langle\hat n^o\rangle)$, where $\hat n^{e(o)}=\sum_{i=\text{even}(\text{odd}),\sigma}\hat n_{i,\sigma}$ ($I$ was measured in Ref.~\cite{Schreiber2015}), and the kinetic energy $K=-t\sum_{i,\sigma}\langle(\hat c^{\dagger}_{i\sigma} \hat c^{}_{i+1,\sigma}+\text{H.c.})\rangle -t'\sum_{i,\sigma}\langle(\hat c_{i\sigma}^\dagger \hat c^{}_{i+2,\sigma}  + \text{H.c.})\rangle$, are directly related to the charge degrees of freedom. The third one, the antiferromagnetic structure factor $S = 1/L\sum_{i,j}e^{i\pi(i-j)}\langle({\hat n_{i\uparrow}}-{\hat n_{i\downarrow}})({\hat n_{j\uparrow}}-{\hat n_{j\downarrow}})\rangle$ is related to the spin degrees of freedom (from now on we refer to it as {\it the structure factor}). The relaxation times of the charge and spin degrees of freedom are expected to be different for very strong interactions \cite{bauer_dorfner_15}.

\begin{figure}[!tb] 
 \includegraphics[width=0.98\columnwidth]{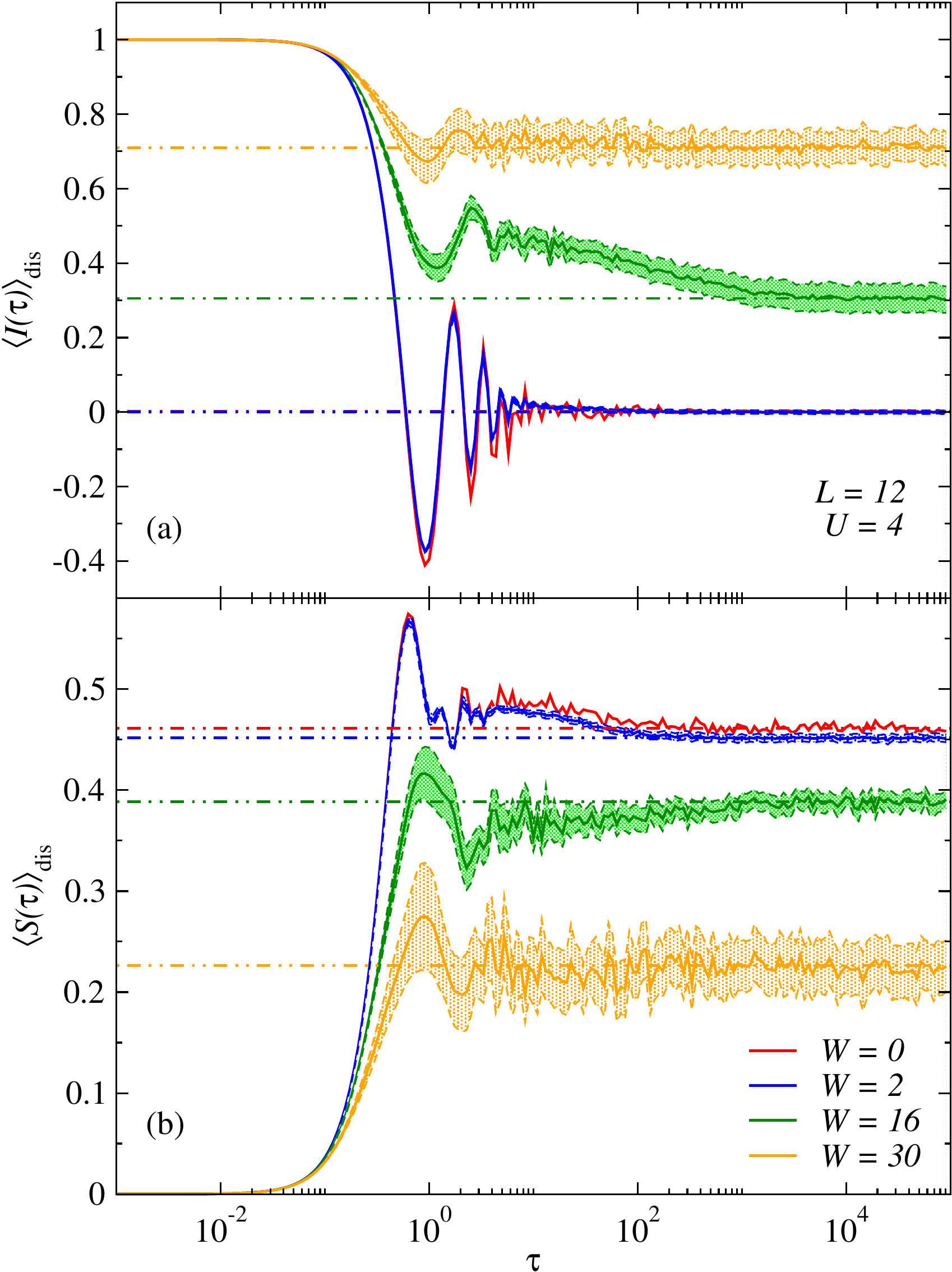}
 \vspace{-0.1cm}
 \caption{(Color online) Disorder averaged results for the time evolution of 
(a) the even-odd site occupation imbalance, and (b) the antiferromagnetic 
structure factor. The shaded area around the curves depicts the standard 
deviation of the mean, after an average over ten disorder
realizations. The horizontal dashed lines depict the disorder averaged values of 
the diagonal ensemble predictions (see text).}
  \label{fig:time_dependence}
\end{figure}

Figure~\ref{fig:time_dependence} displays the disorder averaged time evolution of the imbalance [Fig.~\ref{fig:time_dependence}(a)] and of the structure factor [Fig.~\ref{fig:time_dependence}(b)]. We also display, as horizontal dashed lines, the disorder average of the diagonal ensemble results. Given an observable $\widehat{\mathcal{O}}$, the diagonal ensemble result (which, in the absence of degeneracies, is the same as the infinite-time average of the observable \cite{Rigol2008}) can be obtained as $\mathcal{O}_\text{DE}=\sum_\alpha |C_\alpha|^2 \mathcal{O}_{\alpha\alpha}$, where $\mathcal{O}_{\alpha\alpha}=\langle\alpha|\widehat{\mathcal{O}}|\alpha\rangle$, $|\alpha\rangle$ are the eigenstates of the Hamiltonian ($\hat{H}|\alpha\rangle=E_\alpha|\alpha\rangle$, $E_\alpha$ are the eigenenergies), and $C_\alpha=\langle\alpha|\psi_I\rangle$. We say that $\widehat{\mathcal{O}}$ equilibrates if it relaxes to $\mathcal{O}_\text{DE}$ and remains close to it at later times. 

\begin{figure}[!tb] 
  \includegraphics[width=0.98\columnwidth]{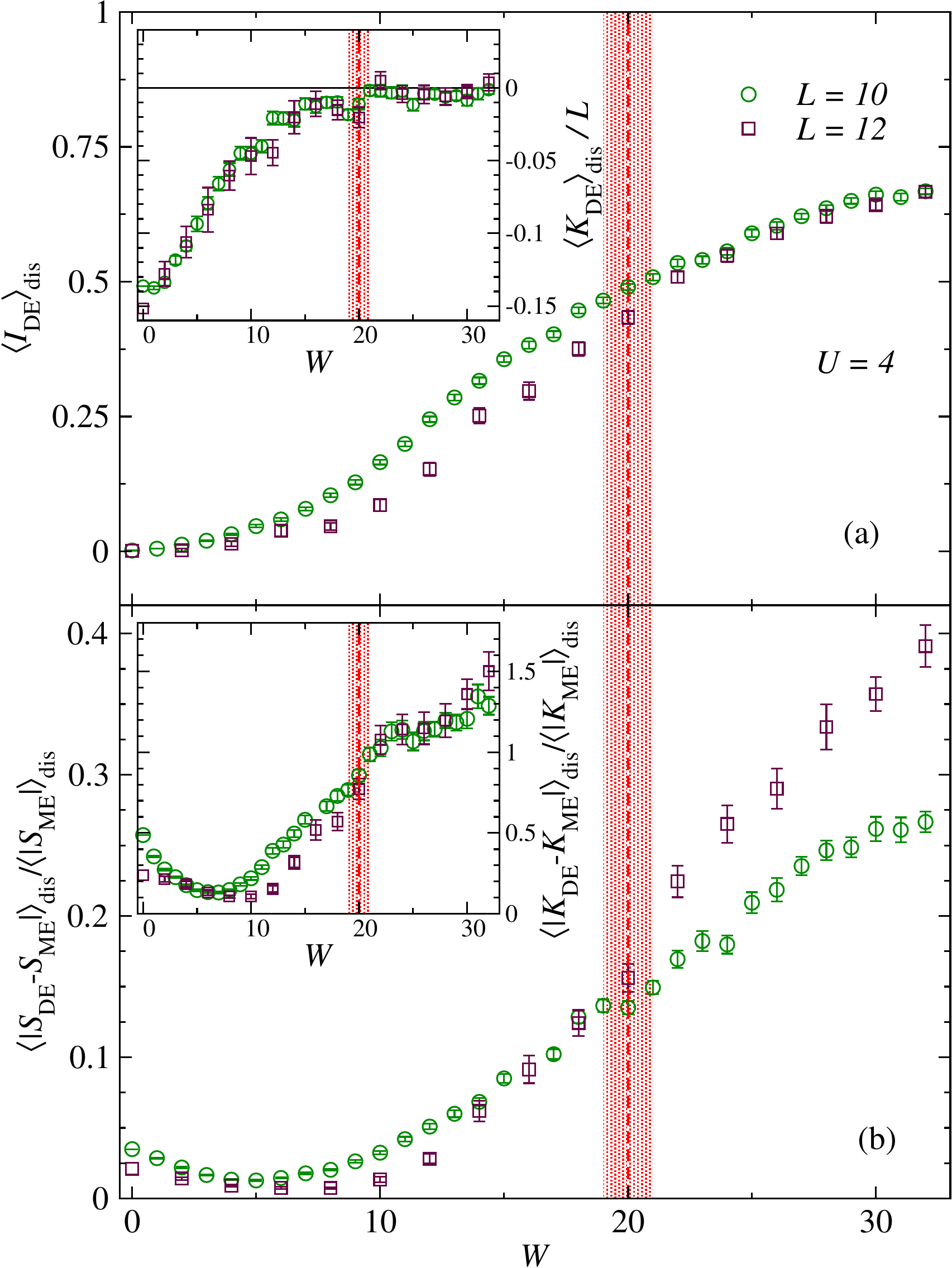}
  \vspace{-0.1cm}
  \caption{(Color online) (a) Disorder averaged diagonal ensemble results for the imbalance (main panel) and the kinetic energy per site (inset) vs the amplitude of the disorder $W$. (b) Normalized disorder average difference between the diagonal and microcanonical ensemble predictions for the structure factor (main panel) and the kinetic energy (inset) vs the amplitude of the disorder $W$. In all cases $U=4$, and the width of the microcanonical energy window is $\Delta E=0.1$. The vertical dashed line marks $W_c$, and the shaded region around it signals the interval of confidence reported in Fig.~\ref{fig:r_w_fixed_tp_v2}(e).}
  \label{fig:diag_ensemble_diff_diag_mic}
\end{figure}

Figures~\ref{fig:time_dependence}(a) and \ref{fig:time_dependence}(b) show that $I$ and $S$ equilibrate in both the delocalized and localized regimes. Sufficiently far away from $W_c$ ($W=0$, 2, and 30 in the figure), one can see that both observables have essentially reached the diagonal ensemble result (or are very close to it) for $\tau\simeq10\,(\hbar/t)$. As the system approaches $W_c(\simeq20$ for $U=4$), we find that equilibration times become much longer. For example, for $W=16$ in Fig.~\ref{fig:r_w_fixed_tp_v2}(a), one can see that $I$ becomes nearly time independent only at $\tau\simeq10^3\,(\hbar/t)$. Very long equilibration times at a delocalization to localization transition have also been observed, for much larger system sizes, in the integrable hard-core boson version of the Aubry-Andr\'e model \cite{gramsch_rigol_12}. Those times represent a challenge for experiments.

Next, we check how the diagonal ensemble results for the observables compare to the microcanonical predictions. Whenever they agree, and equilibration occurs (as we have checked), we say that the system thermalizes. We first consider $I$. Since there is no distinction between even and odd sites in the Hamiltonian, the disorder average of $I$ is expected to be zero in the microcanonical ensemble ($\langle{I}_\text{ME}\rangle_\text{dis}=0$). Hence, as argued in Ref.~\cite{Schreiber2015}, the disorder average of the diagonal ensemble result for $I$ ($\langle{I}_\text{DE}\rangle_\text{dis}$) can be taken to be the order parameter for the MBL phase (it can only differ from zero if the system does not thermalize). Figure~\ref{fig:diag_ensemble_diff_diag_mic}(a) shows $\langle{I}_\text{DE}\rangle_\text{dis}$ vs $W$ for two system sizes. For the (small) system sizes that we can study, $\langle{I}_\text{DE}\rangle_\text{dis}$ can be seen to smoothly increase from zero with increasing $W$. However, comparing the results for the two system sizes, one can see that in the delocalized side (and close to $W_c$ in the MBL side) $\langle{I}_\text{DE}\rangle_\text{dis}$ decreases with increasing system size. This is consistent with the expectation that, in the thermodynamic limit, it will vanish in the delocalized side.

Another order parameter that could be used to locate the MBL transition in experiments is the kinetic energy. As discussed in Ref.~\cite{tang_iyer_15}, the dynamics of one-particle correlations in the MBL phase is quantitatively similar to that in the atomic limit (even if the system is not close to that limit). This means that, in the Heisenberg representation, $\hat{c}_{i,\sigma}^\dagger(\tau)\hat{c}_{j,\sigma}^{}(\tau)\approx
\exp[i(\varepsilon_{i}-\varepsilon_{j})\tau/\hbar]\hat{c}_{i,\sigma}^\dagger(0)\hat{c}_{j,\sigma}^{}(0)$. Given our initial state, that implies that $\langle K_\text{DE}\rangle_\text{dis}\approx0$. In the inset in Fig.~\ref{fig:diag_ensemble_diff_diag_mic}(a), one can see that, indeed, $\langle K_\text{DE}\rangle_\text{dis}\approx0$ for $W\gtrsim W_c$. In Fig.~\ref{fig:diag_ensemble_diff_diag_mic}(b), the inset shows $\langle|K_\text{DE}-K_\text{ME}|\rangle_\text{dis}/\langle|K_\text{ME}|\rangle_\text{dis}$ and the main panel shows $\langle|S_\text{DE}-S_\text{ME}|\rangle_\text{dis}/\langle |S_\text{ME}|\rangle_\text{dis}$. Both normalized differences can be seen to decrease with increasing system size in the delocalized phase and not in the MBL. This is consistent with the expectation that thermalization occurs only in the former.

\begin{figure}[!tb] 
  \includegraphics[width=0.98\columnwidth]{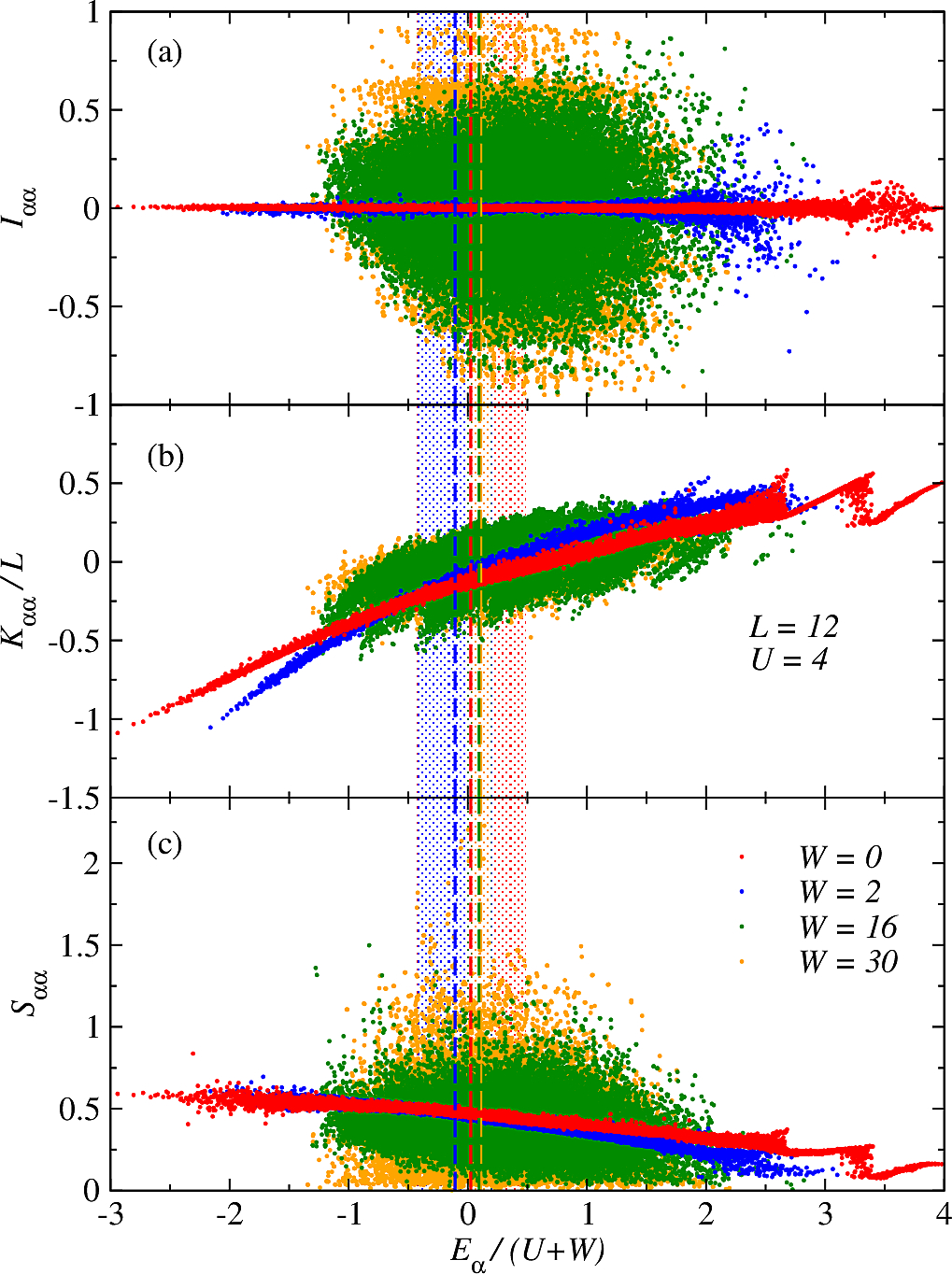}
  \vspace{-0.1cm}
  \caption{(Color online) Eigenstate expectation values of the imbalance (a), the kinetic energy per site (b), and the structure factor (c) for a single disorder realization in systems with four different disorder strengths. The vertical dashed lines show the averaged mean energy, and the shaded are for $W=0$ and 2 and depict the averaged energy width, for the quenches involving this disorder realization. The results reported here were obtained in systems with $U=4$ and $L=12$.}
  \label{fig:obs_spectrum}
\end{figure}

The fact that the system thermalizes (fails to thermalize) in the delocalized (MBL) regime can be understood to be the result of eigenstate thermalization occurring (not occurring) in that regime \cite{Deutsch1991,Srednicki1994,Rigol2008}. In Fig.~\ref{fig:obs_spectrum}, we show the eigenstate expectation values of the three observables of interest for a single disorder realization for different values of $W$. Deep in the delocalized phase ($W=0$ and 2 in the figure), the support for those expectation values at a given energy can be seen to be very small (it decreases with system size, not shown), i.e., eigenstate thermalization occurs. The support of the eigenstate expectation values exhibits a different behavior within the MBL phase, or close to it ($W=16$ and 30 in the figure), i.e., eigenstate thermalization does not occur, or at least, it is not apparent for the system sizes studied. In Fig.~\ref{fig:obs_spectrum}, vertical dashed lines depict the mean energy, and shaded areas around them depict the width of the energy distribution (for $W=0$ and 2), in the quenches involving that disorder realization. They show which part of the spectrum is relevant to the dynamics studied.

\paragraph{Summary and discussion.}
We have studied the ergodic to MBL transition in Hubbard chains. Our main result from the analysis of quantum chaos indicators is that ergodicity is very robust against disorder. Even for on-site interactions as weak as $U=0.2$, we find that the disorder strength required to localize the system is or the order of twice the single-particle bandwidth. We studied the dynamics of those systems starting from a state of the form $|\psi_I\rangle=|\uparrow0\downarrow0\uparrow0\downarrow\ldots\rangle$. We find that various experimentally relevant observables equilibrate in time scales $\sim1-10(\hbar/t)$, whenever the system is not close to the MBL transition. Close to the MBL transition, equilibration times become orders of magnitude longer and might be difficult to reach experimentally. We have also studied the differences between observables after equilibration and the predictions of the microcanonical ensemble finding that, for the small lattice sizes that we are able to study ($\sim12$ sites), they increase smoothly as one increases disorder and can be large even far from the MBL transition. This is reminiscent of the behavior observed as one approaches an integrable point in finite systems \cite{rigol_09a,rigol_09b}. Hence, the analysis of a few small system sizes does not allow one to identify the critical disorder strength at which MBL occurs. Large system sizes, or a careful finite size scaling analysis, are needed. While that might be possible in experiments, it remains a challenge for numerical simulations. Numerical linked cluster expansions \cite{rigol_14,tang_iyer_15}, which exhibit an exponentially fast convergence with increasing the size of the systems that need to be diagonalized \cite{iyer_srednicki_15}, offer a promising way to address this challenge \cite{tang_iyer_15}.

\paragraph{Acknowledgments}
This work was supported by the National Science Foundation Grant No.~PHY13-18303, the US Army Research Office, and CNPq (R.M.). The computations were performed in the Institute for CyberScience at Penn State, the Center for High-Performance Computing at the University of Southern California, and CENAPAD-SP.

\newpage

\onecolumngrid

\begin{center}

\vspace{0.1cm}

{\large \bf Supplementary Materials:
\\ Many-body localization and thermalization in disordered Hubbard chains}\\

\vspace{0.3cm}

\end{center}

\vspace{0.6cm}

\twocolumngrid

\beginsupplement

\section{Lattice sizes and their Hilbert space}
In Table~\ref{table:config}, we show the Hilbert spaces for the different lattice sizes that we can study at quarter filling and at half-filling. At quarter-filling, the Hilbert spaces for $L=6$ and $L=8$ are very small and, as shown in the next section, exhibit erratic behavior. Therefore, in the main text we have focused on the lattices with $L=10$ and 12.

\begin{table}[!ht]
\caption{Lattice configurations and size of the Hilbert space $(D)$ used in this 
work.} 
\centering 
\begin{tabular}{p{0.25\linewidth}p{0.25\linewidth}p{0.25\linewidth}r} 
\hline\hline 
\setlength{\tabcolsep}{35pt}
$L$\quad & $n_{f}$ &$N_\uparrow-N_\downarrow$ &$D$\\ [0.5ex] 
\hline 
6 & 3 & $\pm1$ & 90 \\ 
8 & 4 & 0 & 784 \\
10 & 5 & $\pm1$ & 5400 \\
12 & 6 & 0 & 48400 \\ 
\\
8 & 8 & 0 & 4900 \\
10 & 10 & 0 & 63504\\[1ex] 
\hline 
\end{tabular}
\label{table:config} 
\end{table}

\section{Symmetry-breaking field}
While the extended Hubbard model in Eq.~(1) is non-integrable for nonzero $t^\prime$, in our lattice geometry (which has open boundary conditions) there are still two symmetries, parity and $SU(2)$ when $N_\uparrow=N_\downarrow$, that need to be removed for the level spacing in the system to exhibit a Wigner-Dyson distribution. This is achieved using the symmetry breaking terms described under $\hat{H}_\text{sb}$.

\begin{figure}[!b] 
\includegraphics[width=1\columnwidth]{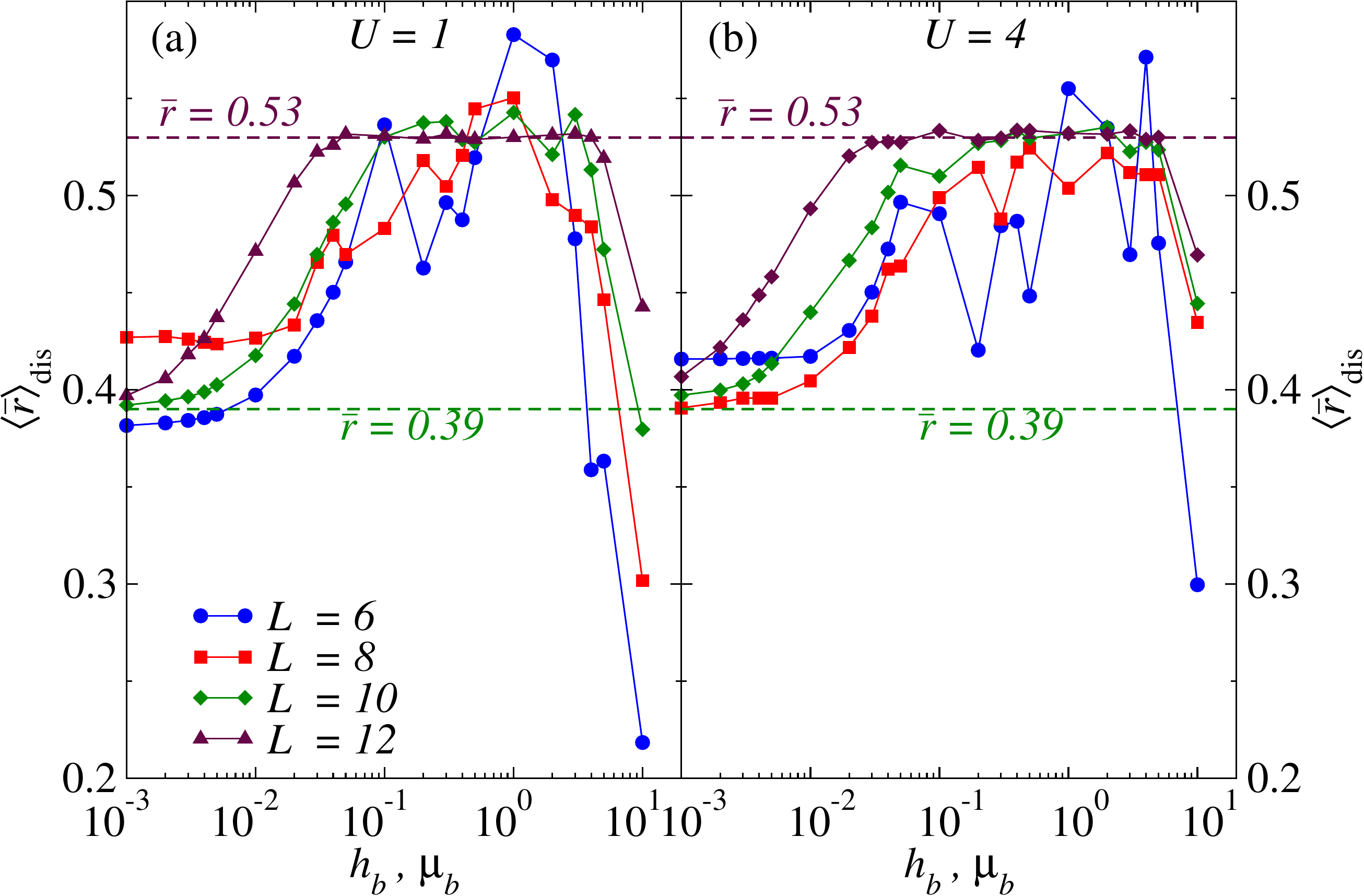}
\caption{(Color online) Average ratio of adjacent energy gaps as a function of the amplitude of the symmetry breaking terms $h_b=\mu_b$ for a chain with $U=1$ (a) and $U=4$ (b), in the absence of disorder. In both cases $t^\prime=0.5$ and the average is computed over the central half of the spectrum.}
\label{fig:hmuborder_analysis}
\end{figure}

Figure~\ref{fig:hmuborder_analysis} shows the dependence of the average ratio of adjacent energy gaps for two values of $U$ and different lattice sizes (in the absence of disorder). For simplicity, we have selected $h_b=\mu_b$, and, as in the main text, $t^\prime = 0.5$. Also, to avoid larger finite size effects related to the low and high energy edges of the spectrum, the averages are calculated over the central half of the spectrum. Due to the smallness of their Hilbert spaces, the systems with $L=6$ and $8$ exhibit very large fluctuations in Fig.~\ref{fig:hmuborder_analysis} (this is the reason they were not shown in the main text). For very small and very large values of $h_b=\mu_b$, one can see that the average ratio of adjacent energy gaps is consistent with that of a Poisson distribution. For small values of $h_b=\mu_b$, the reason are the symmetries mentioned above while, for large values of $h_b=\mu_b$, the reason is that the sites with the fields decouple from the rest of the chain. It is apparent in Fig.~\ref{fig:hmuborder_analysis} that, as the system size increases, the strength of the symmetry breaking terms required to produce a distribution of level spacings that is consistent with the Wigner-Dyson distribution decreases. Hence, for thermodynamically large systems, arbitrary weak symmetry breaking fields would be required.

For the results reported in the main text, we use values of $h_b,\mu_b$ in the range $[0.1,0.5]$. The largest values of the breaking field were necessary in the limit of weak on-site interactions, as the system approached the noninteracting (integrable) limit.

\section{Quasi-periodic potential}

\begin{figure}[!b]
\includegraphics[width=1\columnwidth]{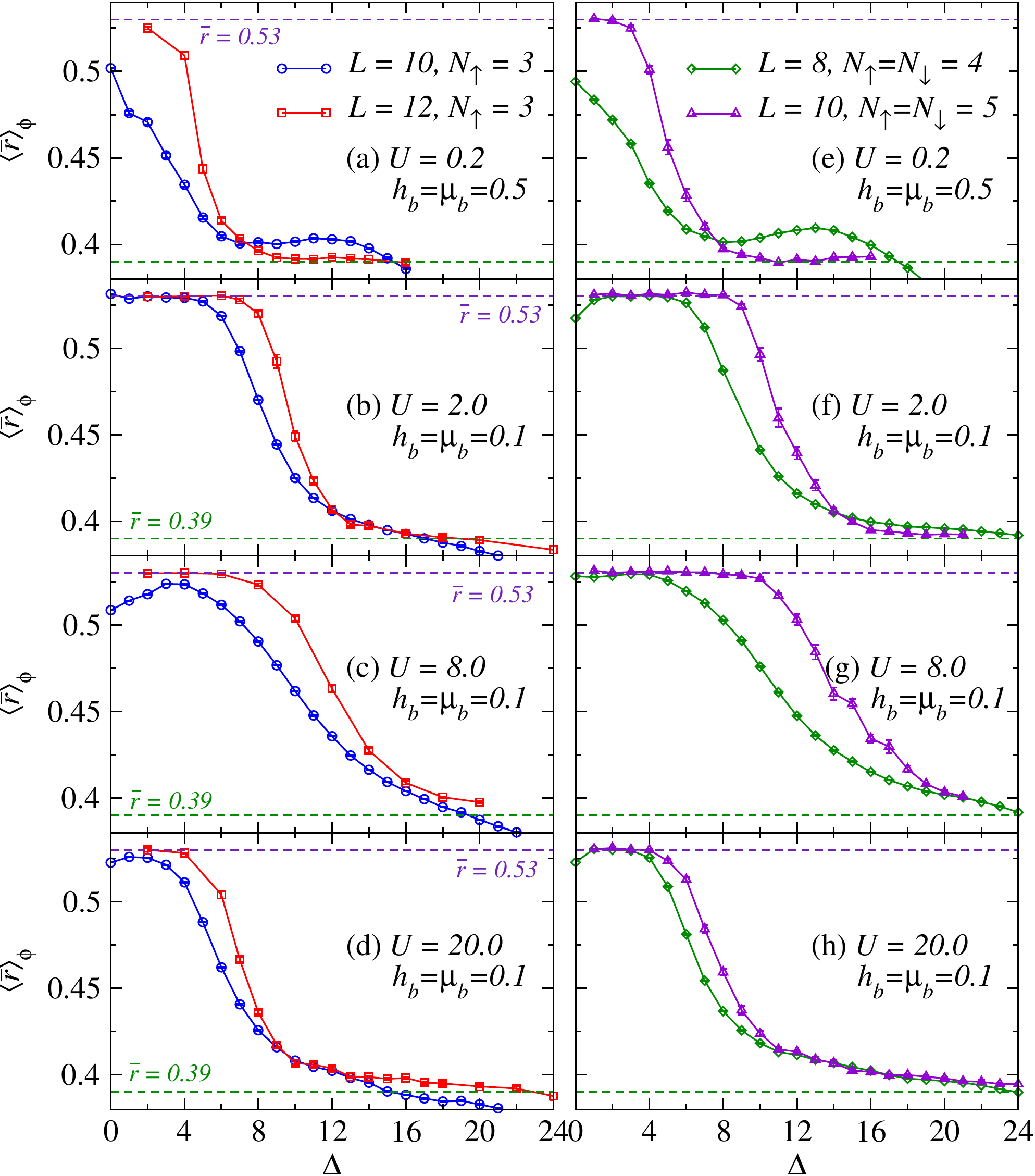}
\caption{(Color online) Average ratio of adjacent energy gaps as a function of the strength of the quasi-periodic potential $\Delta$, for $U=4$, in systems at quarter filling (a) and half filling (b). In both cases, $t^\prime=0.5$, $\bar{r}$ is computed over the central half of the spectrum, and $\langle \cdot \rangle_\phi$ is an average computed over random phases $\phi$.}
\label{fig:aubry_andre_level_spacing}
\end{figure}

We have also studied the MBL transition in the presence of a quasi-periodic potential with $\varepsilon_{i}=\frac{\Delta}{2}\cos{(2\pi\beta i+\phi)}$, where $\Delta$ is the potential strength, $\beta=(\sqrt{5}+1)/2$ is the golden ratio, and $\phi$ is an arbitrary phase that we use to generate different realizations of \{$\varepsilon_{i}$\} (we take $\phi\in [-\pi,\pi]$ with a uniform distribution). For this quasi-periodic potential, unlike for the Anderson case, localization in the noninteracting limit occurs only for $\Delta\geq4$ (in units of $t=1$). As for the disordered case, the phase diagram for this potential reported in Fig.~1(e) was obtained via an analysis of the average ratio of adjacent energy gaps (see Fig.~\ref{fig:aubry_andre_level_spacing}). This phase diagram was explored experimentally in a recent optical lattice experiment~\cite{Schreiber2015}.

Figure~\ref{fig:aubry_andre_level_spacing} reports the equivalent of Fig.~1(a)--1(d) in the main text but for the quasi-periodic potential when the system is at quarter filling [(a)--(d)] and at half-filling [(e)--(h)]. In the latter case, the value of $\Delta$ required to drive the MBL transition is slightly larger in comparison to the former case. This is understandable as $U$ plays an increasingly important role with increasing the filling. We expect a similar behavior to occur in the presence of disorder. Namely, $W_c$ at half-filling should be larger than at quarter filling, which was the case discussed in the main text.

\section{Other quantities supporting MBL}

\begin{figure}[!t] 
 \includegraphics[width=0.98\columnwidth]{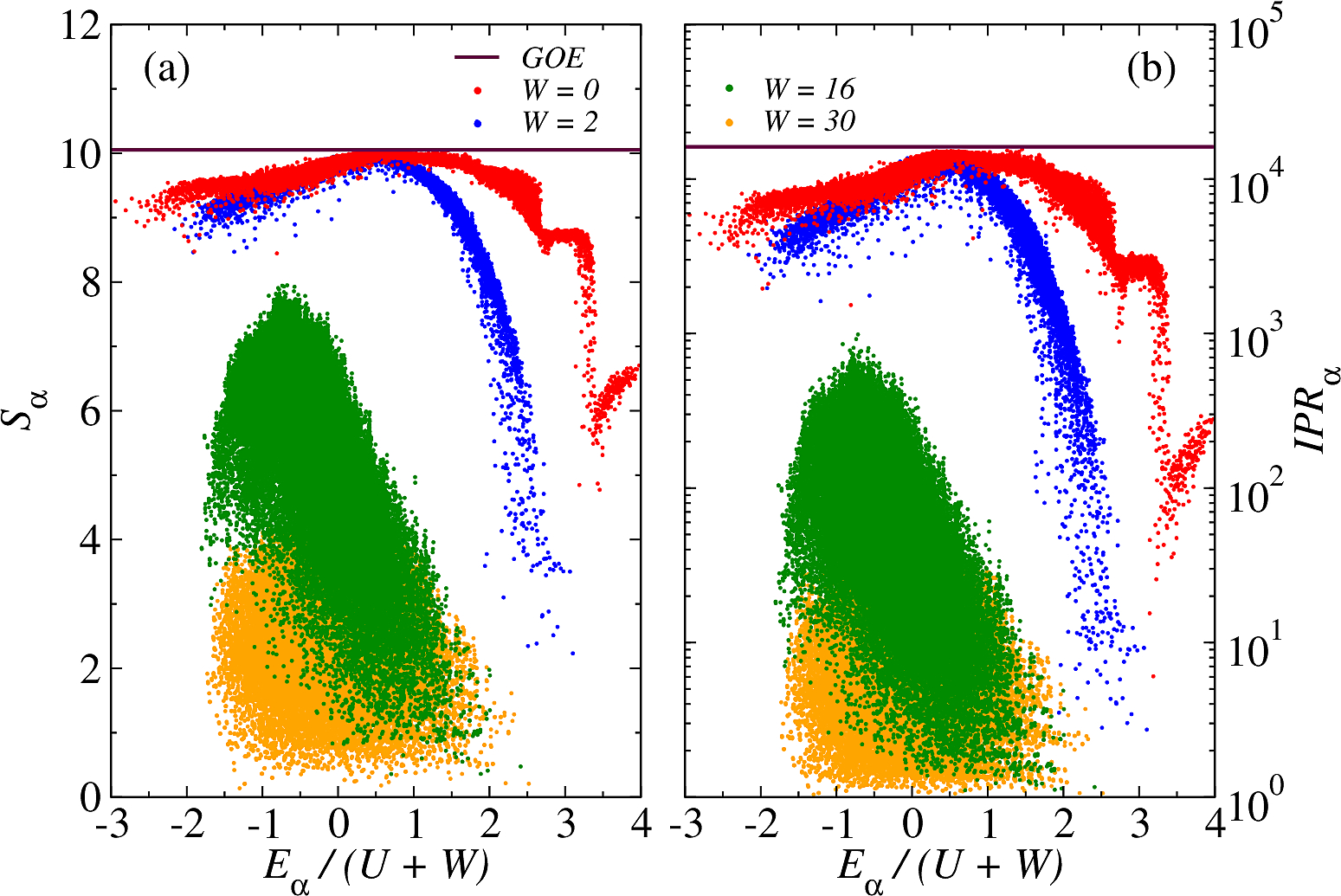}
 \vspace{-0.1cm}
 \caption{(Color online) (a) Shannon entropy of the eigenstates of the Hamiltonian in the computational (Fock) basis vs the eigenstate energies. (b) Same as (a) for the inverse participation ratio. The parameters are: $L=12,\, U = 4,$ and $t^\prime=0.5$. Results are presented for a single realization of disorder and four values of $W$. The horizontal lines depict the GOE predictions.}
  \label{fig:entropy_ipr}
\end{figure}

Other quantum chaos indicators that can be used to locate the ergodic to MBL transition are the Shannon entropy, also known as the information entropy,
\begin{equation}
 S_\alpha \equiv 
-\left(\sum_{j=1}^{D} |c_\alpha^j|^2\ln|c_\alpha^j|^2\right),
\end{equation}
and the inverse participation ratio (IPR),
\begin{equation}
 {\rm IPR}_\alpha = \frac{1}{\sum_{j=1}^{D}|c_\alpha^j|^4},
\end{equation}
where $D$ is the dimension of the Hilbert space, and the coefficients $c_\alpha^j$ correspond to the $j$-th component of the energy eigenvector $|\alpha\rangle$ in some basis. Here, we use the computational (Fock) basis. As one increases the system size in the ergodic (chaotic) regime, one expects $S_\alpha$ to approach the Gaussian orthogonal ensemble (GOE) prediction $S_\alpha^{\rm GOE}=\ln(0.48 D)$, provided $|\alpha\rangle$ is away from the edges of the spectrum \cite{santos_rigol_10a}. For ${\rm IPR}_\alpha$ one expects that it should approach ${\rm IPR}^{\rm{GOE}} = (D+2)/3$. Figure \ref{fig:entropy_ipr} shows that, when $W=0$ and 2 for $L=12$, both $S_\alpha$ and ${\rm IPR}_\alpha$ are indeed close to the GOE predictions away from the edges of the spectrum. For $W=16$ and 30, on the other hand, localization in the Fock basis is made apparent by the small values of $S_\alpha$ and ${\rm IPR}_\alpha$ over the entire spectrum.

\begin{figure}[!t] 
  \includegraphics[width=0.98\columnwidth]{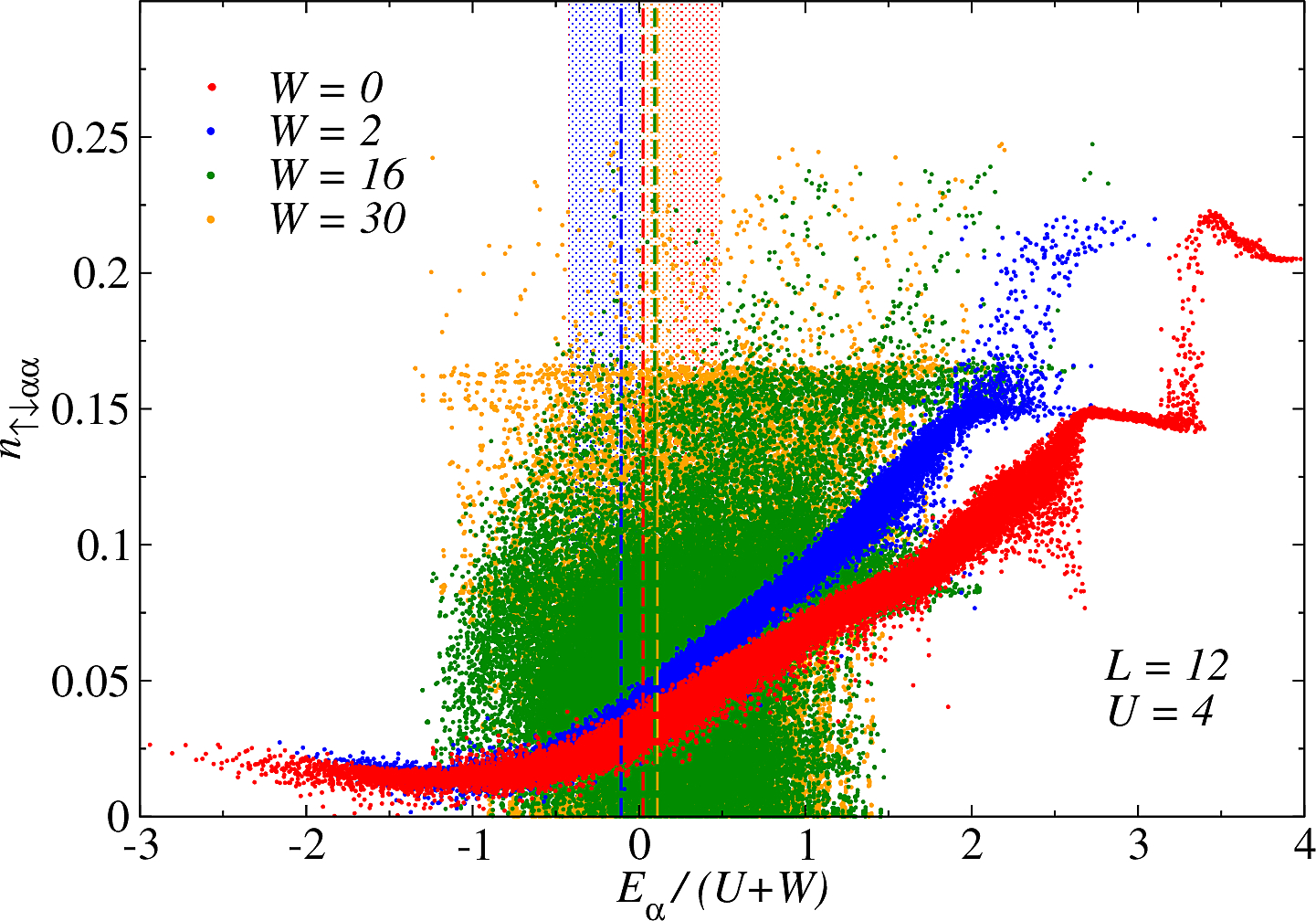}
  \vspace{-0.1cm}
  \caption{(Color online) Same as Fig.~4 in the main text but for the double occupancy.}
  \label{fig:obs_spectrum_suppl}
\end{figure}

In Fig.~\ref{fig:obs_spectrum_suppl}, we show the energy eigenstate expectation values of another observable that is of interest to experiments with ultracold fermions in optical lattices, namely, the double occupancy: $\hat n_{\uparrow\downarrow} = 1/L\sum_i({\hat n_{i\uparrow}}{\hat n_{i\downarrow}})$. The results for this observable are qualitatively similar to those discussed in the main text (see Fig.~4 there) for the imbalance, the kinetic energy, and the structure factor.

\newpage

\begin{thebibliography}{39}%
\makeatletter
\providecommand \@ifxundefined [1]{%
 \@ifx{#1\undefined}
}%
\providecommand \@ifnum [1]{%
 \ifnum #1\expandafter \@firstoftwo
 \else \expandafter \@secondoftwo
 \fi
}%
\providecommand \@ifx [1]{%
 \ifx #1\expandafter \@firstoftwo
 \else \expandafter \@secondoftwo
 \fi
}%
\providecommand \natexlab [1]{#1}%
\providecommand \enquote  [1]{``#1''}%
\providecommand \bibnamefont  [1]{#1}%
\providecommand \bibfnamefont [1]{#1}%
\providecommand \citenamefont [1]{#1}%
\providecommand \href@noop [0]{\@secondoftwo}%
\providecommand \href [0]{\begingroup \@sanitize@url \@href}%
\providecommand \@href[1]{\@@startlink{#1}\@@href}%
\providecommand \@@href[1]{\endgroup#1\@@endlink}%
\providecommand \@sanitize@url [0]{\catcode `\\12\catcode `\$12\catcode
  `\&12\catcode `\#12\catcode `\^12\catcode `\_12\catcode `\%12\relax}%
\providecommand \@@startlink[1]{}%
\providecommand \@@endlink[0]{}%
\providecommand \url  [0]{\begingroup\@sanitize@url \@url }%
\providecommand \@url [1]{\endgroup\@href {#1}{\urlprefix }}%
\providecommand \urlprefix  [0]{URL }%
\providecommand \Eprint [0]{\href }%
\providecommand \doibase [0]{http://dx.doi.org/}%
\providecommand \selectlanguage [0]{\@gobble}%
\providecommand \bibinfo  [0]{\@secondoftwo}%
\providecommand \bibfield  [0]{\@secondoftwo}%
\providecommand \translation [1]{[#1]}%
\providecommand \BibitemOpen [0]{}%
\providecommand \bibitemStop [0]{}%
\providecommand \bibitemNoStop [0]{.\EOS\space}%
\providecommand \EOS [0]{\spacefactor3000\relax}%
\providecommand \BibitemShut  [1]{\csname bibitem#1\endcsname}%
\let\auto@bib@innerbib\@empty
\bibitem [{\citenamefont {Anderson}(1958)}]{Anderson1958}%
  \BibitemOpen
  \bibfield  {author} {\bibinfo {author} {\bibfnamefont {P.~W.}\ \bibnamefont
  {Anderson}},\ }\bibfield  {title} {\enquote {\bibinfo {title} {Absence of
  diffusion in certain random lattices},}\ }\href@noop {} {\bibfield  {journal}
  {\bibinfo  {journal} {Phys. Rev.}\ }\textbf {\bibinfo {volume} {109}},\
  \bibinfo {pages} {1492} (\bibinfo {year} {1958})}\BibitemShut {NoStop}%
\bibitem [{\citenamefont {Abrahams}\ \emph {et~al.}(1979)\citenamefont
  {Abrahams}, \citenamefont {Anderson}, \citenamefont {Licciardello},\ and\
  \citenamefont {Ramakrishnan}}]{Abrahams1979}%
  \BibitemOpen
  \bibfield  {author} {\bibinfo {author} {\bibfnamefont {E.}~\bibnamefont
  {Abrahams}}, \bibinfo {author} {\bibfnamefont {P.~W.}\ \bibnamefont
  {Anderson}}, \bibinfo {author} {\bibfnamefont {D.~C.}\ \bibnamefont
  {Licciardello}}, \ and\ \bibinfo {author} {\bibfnamefont {T.~V.}\
  \bibnamefont {Ramakrishnan}},\ }\bibfield  {title} {\enquote {\bibinfo
  {title} {Scaling theory of localization: Absence of quantum diffusion in two
  dimensions},}\ }\href@noop {} {\bibfield  {journal} {\bibinfo  {journal}
  {Phys. Rev. Lett.}\ }\textbf {\bibinfo {volume} {42}},\ \bibinfo {pages}
  {673--676} (\bibinfo {year} {1979})}\BibitemShut {NoStop}%
\bibitem [{\citenamefont {Fleishman}\ and\ \citenamefont
  {Anderson}(1980)}]{fleishman80}%
  \BibitemOpen
  \bibfield  {author} {\bibinfo {author} {\bibfnamefont {L.}~\bibnamefont
  {Fleishman}}\ and\ \bibinfo {author} {\bibfnamefont {P.~W.}\ \bibnamefont
  {Anderson}},\ }\bibfield  {title} {\enquote {\bibinfo {title} {Interactions
  and the {A}nderson transition},}\ }\href {\doibase 10.1103/PhysRevB.21.2366}
  {\bibfield  {journal} {\bibinfo  {journal} {Phys. Rev. B}\ }\textbf {\bibinfo
  {volume} {21}},\ \bibinfo {pages} {2366--2377} (\bibinfo {year}
  {1980})}\BibitemShut {NoStop}%
\bibitem [{\citenamefont {Altshuler}\ \emph {et~al.}(1997)\citenamefont
  {Altshuler}, \citenamefont {Gefen}, \citenamefont {Kamenev},\ and\
  \citenamefont {Levitov}}]{altshuler_gefen_97}%
  \BibitemOpen
  \bibfield  {author} {\bibinfo {author} {\bibfnamefont {B.~L.}\ \bibnamefont
  {Altshuler}}, \bibinfo {author} {\bibfnamefont {Y.}~\bibnamefont {Gefen}},
  \bibinfo {author} {\bibfnamefont {A.}~\bibnamefont {Kamenev}}, \ and\
  \bibinfo {author} {\bibfnamefont {L.~S.}\ \bibnamefont {Levitov}},\
  }\bibfield  {title} {\enquote {\bibinfo {title} {Quasiparticle lifetime in a
  finite system: {A} nonperturbative approach},}\ }\href {\doibase
  10.1103/PhysRevLett.78.2803} {\bibfield  {journal} {\bibinfo  {journal}
  {Phys. Rev. Lett.}\ }\textbf {\bibinfo {volume} {78}},\ \bibinfo {pages}
  {2803--2806} (\bibinfo {year} {1997})}\BibitemShut {NoStop}%
\bibitem [{\citenamefont {Gornyi}\ \emph {et~al.}(2005)\citenamefont {Gornyi},
  \citenamefont {Mirlin},\ and\ \citenamefont {Polyakov}}]{gornyi_mirlin_05}%
  \BibitemOpen
  \bibfield  {author} {\bibinfo {author} {\bibfnamefont {I.~V.}\ \bibnamefont
  {Gornyi}}, \bibinfo {author} {\bibfnamefont {A.~D.}\ \bibnamefont {Mirlin}},
  \ and\ \bibinfo {author} {\bibfnamefont {D.~G.}\ \bibnamefont {Polyakov}},\
  }\bibfield  {title} {\enquote {\bibinfo {title} {Interacting electrons in
  disordered wires: {A}nderson localization and low-${T}$ transport},}\ }\href
  {\doibase 10.1103/PhysRevLett.95.206603} {\bibfield  {journal} {\bibinfo
  {journal} {Phys. Rev. Lett.}\ }\textbf {\bibinfo {volume} {95}},\ \bibinfo
  {pages} {206603} (\bibinfo {year} {2005})}\BibitemShut {NoStop}%
\bibitem [{\citenamefont {Basko}\ \emph {et~al.}(2006)\citenamefont {Basko},
  \citenamefont {Aleiner},\ and\ \citenamefont {Altshuler}}]{basko06}%
  \BibitemOpen
  \bibfield  {author} {\bibinfo {author} {\bibfnamefont {D.~M.}\ \bibnamefont
  {Basko}}, \bibinfo {author} {\bibfnamefont {I.~L.}\ \bibnamefont {Aleiner}},
  \ and\ \bibinfo {author} {\bibfnamefont {B.~L.}\ \bibnamefont {Altshuler}},\
  }\bibfield  {title} {\enquote {\bibinfo {title} {Metal–insulator transition
  in a weakly interacting many-electron system with localized single-particle
  states},}\ }\href {\doibase http://dx.doi.org/10.1016/j.aop.2005.11.014}
  {\bibfield  {journal} {\bibinfo  {journal} {Ann. Phys.}\ }\textbf {\bibinfo
  {volume} {321}},\ \bibinfo {pages} {1126 -- 1205} (\bibinfo {year}
  {2006})}\BibitemShut {NoStop}%
\bibitem [{\citenamefont {Oganesyan}\ and\ \citenamefont
  {Huse}(2007)}]{oganesyan_huse_07}%
  \BibitemOpen
  \bibfield  {author} {\bibinfo {author} {\bibfnamefont {V.}~\bibnamefont
  {Oganesyan}}\ and\ \bibinfo {author} {\bibfnamefont {D.~A.}\ \bibnamefont
  {Huse}},\ }\bibfield  {title} {\enquote {\bibinfo {title} {Localization of
  interacting fermions at high temperature},}\ }\href {\doibase
  10.1103/PhysRevB.75.155111} {\bibfield  {journal} {\bibinfo  {journal} {Phys.
  Rev. B}\ }\textbf {\bibinfo {volume} {75}},\ \bibinfo {pages} {155111}
  (\bibinfo {year} {2007})}\BibitemShut {NoStop}%
\bibitem [{\citenamefont {\ifmmode \check{Z}\else
  \v{Z}\fi{}nidari\ifmmode~\check{c}\else \v{c}\fi{}}\ \emph
  {et~al.}(2008)\citenamefont {\ifmmode \check{Z}\else
  \v{Z}\fi{}nidari\ifmmode~\check{c}\else \v{c}\fi{}}, \citenamefont {Prosen},\
  and\ \citenamefont {Prelov\ifmmode~\check{s}\else
  \v{s}\fi{}ek}}]{znidaric08}%
  \BibitemOpen
  \bibfield  {author} {\bibinfo {author} {\bibfnamefont {M.}~\bibnamefont
  {\ifmmode \check{Z}\else \v{Z}\fi{}nidari\ifmmode~\check{c}\else
  \v{c}\fi{}}}, \bibinfo {author} {\bibfnamefont {T.}~\bibnamefont {Prosen}}, \
  and\ \bibinfo {author} {\bibfnamefont {P.}~\bibnamefont
  {Prelov\ifmmode~\check{s}\else \v{s}\fi{}ek}},\ }\bibfield  {title} {\enquote
  {\bibinfo {title} {Many-body localization in the heisenberg ${XXZ}$ magnet in
  a random field},}\ }\href {\doibase 10.1103/PhysRevB.77.064426} {\bibfield
  {journal} {\bibinfo  {journal} {Phys. Rev. B}\ }\textbf {\bibinfo {volume}
  {77}},\ \bibinfo {pages} {064426} (\bibinfo {year} {2008})}\BibitemShut
  {NoStop}%
\bibitem [{\citenamefont {Pal}\ and\ \citenamefont {Huse}(2010)}]{pal10}%
  \BibitemOpen
  \bibfield  {author} {\bibinfo {author} {\bibfnamefont {A.}~\bibnamefont
  {Pal}}\ and\ \bibinfo {author} {\bibfnamefont {D.~A.}\ \bibnamefont {Huse}},\
  }\bibfield  {title} {\enquote {\bibinfo {title} {Many-body localization phase
  transition},}\ }\href {\doibase 10.1103/PhysRevB.82.174411} {\bibfield
  {journal} {\bibinfo  {journal} {Phys. Rev. B}\ }\textbf {\bibinfo {volume}
  {82}},\ \bibinfo {pages} {174411} (\bibinfo {year} {2010})}\BibitemShut
  {NoStop}%
\bibitem [{\citenamefont {Khatami}\ \emph {et~al.}(2012)\citenamefont
  {Khatami}, \citenamefont {Rigol}, \citenamefont {Rela\~no},\ and\
  \citenamefont {Garcia-Garcia}}]{khatami_rigol_12}%
  \BibitemOpen
  \bibfield  {author} {\bibinfo {author} {\bibfnamefont {E.}~\bibnamefont
  {Khatami}}, \bibinfo {author} {\bibfnamefont {M.}~\bibnamefont {Rigol}},
  \bibinfo {author} {\bibfnamefont {A.}~\bibnamefont {Rela\~no}}, \ and\
  \bibinfo {author} {\bibfnamefont {A.~M.}\ \bibnamefont {Garcia-Garcia}},\
  }\bibfield  {title} {\enquote {\bibinfo {title} {Quantum quenches in
  disordered systems: Approach to thermal equilibrium without a typical
  relaxation time},}\ }\href {\doibase 10.1103/PhysRevE.85.050102} {\bibfield
  {journal} {\bibinfo  {journal} {Phys. Rev. E}\ }\textbf {\bibinfo {volume}
  {85}},\ \bibinfo {pages} {050102(R)} (\bibinfo {year} {2012})}\BibitemShut
  {NoStop}%
\bibitem [{\citenamefont {Bardarson}\ \emph {et~al.}(2012)\citenamefont
  {Bardarson}, \citenamefont {Pollmann},\ and\ \citenamefont
  {Moore}}]{bardarson_pollmann_12}%
  \BibitemOpen
  \bibfield  {author} {\bibinfo {author} {\bibfnamefont {J.~H.}\ \bibnamefont
  {Bardarson}}, \bibinfo {author} {\bibfnamefont {F.}~\bibnamefont {Pollmann}},
  \ and\ \bibinfo {author} {\bibfnamefont {J.~E.}\ \bibnamefont {Moore}},\
  }\bibfield  {title} {\enquote {\bibinfo {title} {Unbounded growth of
  entanglement in models of many-body localization},}\ }\href {\doibase
  10.1103/PhysRevLett.109.017202} {\bibfield  {journal} {\bibinfo  {journal}
  {Phys. Rev. Lett.}\ }\textbf {\bibinfo {volume} {109}},\ \bibinfo {pages}
  {017202} (\bibinfo {year} {2012})}\BibitemShut {NoStop}%
\bibitem [{\citenamefont {Nandkishore}\ and\ \citenamefont
  {Huse}(2015)}]{nandkishore_huse_review_15}%
  \BibitemOpen
  \bibfield  {author} {\bibinfo {author} {\bibfnamefont {R.}~\bibnamefont
  {Nandkishore}}\ and\ \bibinfo {author} {\bibfnamefont {D.~A.}\ \bibnamefont
  {Huse}},\ }\bibfield  {title} {\enquote {\bibinfo {title} {Many-body
  localization and thermalization in quantum statistical mechanics},}\
  }\href@noop {} {\bibfield  {journal} {\bibinfo  {journal} {Annual Review of
  Condensed Matter Physics}\ }\textbf {\bibinfo {volume} {6}},\ \bibinfo
  {pages} {15--38} (\bibinfo {year} {2015})}\BibitemShut {NoStop}%
\bibitem [{\citenamefont {Altman}\ and\ \citenamefont
  {Vosk}(2015)}]{altman_vosk_review_15}%
  \BibitemOpen
  \bibfield  {author} {\bibinfo {author} {\bibfnamefont {E.}~\bibnamefont
  {Altman}}\ and\ \bibinfo {author} {\bibfnamefont {R.}~\bibnamefont {Vosk}},\
  }\bibfield  {title} {\enquote {\bibinfo {title} {Universal dynamics and
  renormalization in many-body-localized systems},}\ }\href@noop {} {\bibfield
  {journal} {\bibinfo  {journal} {Annual Review of Condensed Matter Physics}\
  }\textbf {\bibinfo {volume} {6}},\ \bibinfo {pages} {383--409} (\bibinfo
  {year} {2015})}\BibitemShut {NoStop}%
\bibitem [{\citenamefont {Kondov}\ \emph {et~al.}(2015)\citenamefont {Kondov},
  \citenamefont {McGehee}, \citenamefont {Xu},\ and\ \citenamefont
  {DeMarco}}]{kondov_mcgehee_15}%
  \BibitemOpen
  \bibfield  {author} {\bibinfo {author} {\bibfnamefont {S.~S.}\ \bibnamefont
  {Kondov}}, \bibinfo {author} {\bibfnamefont {W.~R.}\ \bibnamefont {McGehee}},
  \bibinfo {author} {\bibfnamefont {W.}~\bibnamefont {Xu}}, \ and\ \bibinfo
  {author} {\bibfnamefont {B.}~\bibnamefont {DeMarco}},\ }\bibfield  {title}
  {\enquote {\bibinfo {title} {Disorder-induced localization in a strongly
  correlated atomic {H}ubbard gas},}\ }\href {\doibase
  10.1103/PhysRevLett.114.083002} {\bibfield  {journal} {\bibinfo  {journal}
  {Phys. Rev. Lett.}\ }\textbf {\bibinfo {volume} {114}},\ \bibinfo {pages}
  {083002} (\bibinfo {year} {2015})}\BibitemShut {NoStop}%
\bibitem [{\citenamefont {Schreiber}\ \emph {et~al.}(2015)\citenamefont
  {Schreiber}, \citenamefont {Hodgman}, \citenamefont {Bordia}, \citenamefont
  {L\"uschen}, \citenamefont {Fischer}, \citenamefont {Vosk}, \citenamefont
  {Altman}, \citenamefont {Schneider},\ and\ \citenamefont
  {Bloch}}]{Schreiber2015}%
  \BibitemOpen
  \bibfield  {author} {\bibinfo {author} {\bibfnamefont {M.}~\bibnamefont
  {Schreiber}}, \bibinfo {author} {\bibfnamefont {S.~S.}\ \bibnamefont
  {Hodgman}}, \bibinfo {author} {\bibfnamefont {P.}~\bibnamefont {Bordia}},
  \bibinfo {author} {\bibfnamefont {H.~P.}\ \bibnamefont {L\"uschen}}, \bibinfo
  {author} {\bibfnamefont {M.~H.}\ \bibnamefont {Fischer}}, \bibinfo {author}
  {\bibfnamefont {R.}~\bibnamefont {Vosk}}, \bibinfo {author} {\bibfnamefont
  {E.}~\bibnamefont {Altman}}, \bibinfo {author} {\bibfnamefont
  {U.}~\bibnamefont {Schneider}}, \ and\ \bibinfo {author} {\bibfnamefont
  {I.}~\bibnamefont {Bloch}},\ }\bibfield  {title} {\enquote {\bibinfo {title}
  {Observation of many-body localization of interacting fermions in a
  quasi-random optical lattice},}\ }\href@noop {} {\bibfield  {journal}
  {\bibinfo  {journal} {Science}\ }\textbf {\bibinfo {volume} {349}},\ \bibinfo
  {pages} {842} (\bibinfo {year} {2015})}\BibitemShut {NoStop}%
\bibitem [{\citenamefont {Bordia}\ \emph {et~al.}(2015)\citenamefont {Bordia},
  \citenamefont {L{\"u}schen}, \citenamefont {Hodgman}, \citenamefont
  {Schreiber}, \citenamefont {Bloch},\ and\ \citenamefont
  {Schneider}}]{Bordia2015}%
  \BibitemOpen
  \bibfield  {author} {\bibinfo {author} {\bibfnamefont {P.}~\bibnamefont
  {Bordia}}, \bibinfo {author} {\bibfnamefont {H.~P.}\ \bibnamefont
  {L{\"u}schen}}, \bibinfo {author} {\bibfnamefont {S.~S.}\ \bibnamefont
  {Hodgman}}, \bibinfo {author} {\bibfnamefont {M.}~\bibnamefont {Schreiber}},
  \bibinfo {author} {\bibfnamefont {I.}~\bibnamefont {Bloch}}, \ and\ \bibinfo
  {author} {\bibfnamefont {U.}~\bibnamefont {Schneider}},\ }\bibfield  {title}
  {\enquote {\bibinfo {title} {Coupling identical {1D} many-body localized
  systems},}\ }\href@noop {} {\bibfield  {journal} {\bibinfo  {journal}
  {arXiv:1509.00478}\ } (\bibinfo {year} {2015})}\BibitemShut {NoStop}%
\bibitem [{\citenamefont {Smith}\ \emph {et~al.}(2015)\citenamefont {Smith},
  \citenamefont {Lee}, \citenamefont {Richerme}, \citenamefont {Neyenhuis},
  \citenamefont {Hess}, \citenamefont {Hauke}, \citenamefont {Heyl},
  \citenamefont {Huse},\ and\ \citenamefont {Monroe}}]{Smith2015}%
  \BibitemOpen
  \bibfield  {author} {\bibinfo {author} {\bibfnamefont {J.}~\bibnamefont
  {Smith}}, \bibinfo {author} {\bibfnamefont {A.}~\bibnamefont {Lee}}, \bibinfo
  {author} {\bibfnamefont {P.}~\bibnamefont {Richerme}}, \bibinfo {author}
  {\bibfnamefont {B.}~\bibnamefont {Neyenhuis}}, \bibinfo {author}
  {\bibfnamefont {P.~W.}\ \bibnamefont {Hess}}, \bibinfo {author}
  {\bibfnamefont {P.}~\bibnamefont {Hauke}}, \bibinfo {author} {\bibfnamefont
  {M.}~\bibnamefont {Heyl}}, \bibinfo {author} {\bibfnamefont {D.~A.}\
  \bibnamefont {Huse}}, \ and\ \bibinfo {author} {\bibfnamefont
  {C.}~\bibnamefont {Monroe}},\ }\bibfield  {title} {\enquote {\bibinfo {title}
  {Many-body localization in a quantum simulator with programmable random
  disorder},}\ }\href@noop {} {\bibfield  {journal} {\bibinfo  {journal}
  {arXiv:1508.07026}\ } (\bibinfo {year} {2015})}\BibitemShut {NoStop}%
\bibitem [{\citenamefont {Deutsch}(1991)}]{Deutsch1991}%
  \BibitemOpen
  \bibfield  {author} {\bibinfo {author} {\bibfnamefont {J.~M.}\ \bibnamefont
  {Deutsch}},\ }\bibfield  {title} {\enquote {\bibinfo {title} {Quantum
  statistical mechanics in a closed system},}\ }\href@noop {} {\bibfield
  {journal} {\bibinfo  {journal} {Phys. Rev. A}\ }\textbf {\bibinfo {volume}
  {43}},\ \bibinfo {pages} {2046} (\bibinfo {year} {1991})}\BibitemShut
  {NoStop}%
\bibitem [{\citenamefont {Srednicki}(1994)}]{Srednicki1994}%
  \BibitemOpen
  \bibfield  {author} {\bibinfo {author} {\bibfnamefont {M.}~\bibnamefont
  {Srednicki}},\ }\bibfield  {title} {\enquote {\bibinfo {title} {Chaos and
  quantum thermalization},}\ }\href@noop {} {\bibfield  {journal} {\bibinfo
  {journal} {Physical Review E}\ }\textbf {\bibinfo {volume} {50}},\ \bibinfo
  {pages} {888} (\bibinfo {year} {1994})}\BibitemShut {NoStop}%
\bibitem [{\citenamefont {Rigol}\ \emph {et~al.}(2008)\citenamefont {Rigol},
  \citenamefont {Dunjko},\ and\ \citenamefont {Olshanii}}]{Rigol2008}%
  \BibitemOpen
  \bibfield  {author} {\bibinfo {author} {\bibfnamefont {M.}~\bibnamefont
  {Rigol}}, \bibinfo {author} {\bibfnamefont {V.}~\bibnamefont {Dunjko}}, \
  and\ \bibinfo {author} {\bibfnamefont {M.}~\bibnamefont {Olshanii}},\
  }\bibfield  {title} {\enquote {\bibinfo {title} {Thermalization and its
  mechanism for generic isolated quantum systems},}\ }\href@noop {} {\bibfield
  {journal} {\bibinfo  {journal} {Nature}\ }\textbf {\bibinfo {volume} {452}},\
  \bibinfo {pages} {854--858} (\bibinfo {year} {2008})}\BibitemShut {NoStop}%
\bibitem [{\citenamefont {Rigol}(2009{\natexlab{a}})}]{rigol_09a}%
  \BibitemOpen
  \bibfield  {author} {\bibinfo {author} {\bibfnamefont {M.}~\bibnamefont
  {Rigol}},\ }\bibfield  {title} {\enquote {\bibinfo {title} {Breakdown of
  thermalization in finite one-dimensional systems},}\ }\href {\doibase
  10.1103/PhysRevLett.103.100403} {\bibfield  {journal} {\bibinfo  {journal}
  {Phys. Rev. Lett.}\ }\textbf {\bibinfo {volume} {103}},\ \bibinfo {pages}
  {100403} (\bibinfo {year} {2009}{\natexlab{a}})}\BibitemShut {NoStop}%
\bibitem [{\citenamefont {Rigol}(2009{\natexlab{b}})}]{rigol_09b}%
  \BibitemOpen
  \bibfield  {author} {\bibinfo {author} {\bibfnamefont {M.}~\bibnamefont
  {Rigol}},\ }\bibfield  {title} {\enquote {\bibinfo {title} {Quantum quenches
  and thermalization in one-dimensional fermionic systems},}\ }\href {\doibase
  10.1103/PhysRevA.80.053607} {\bibfield  {journal} {\bibinfo  {journal} {Phys.
  Rev. A}\ }\textbf {\bibinfo {volume} {80}},\ \bibinfo {pages} {053607}
  (\bibinfo {year} {2009}{\natexlab{b}})}\BibitemShut {NoStop}%
\bibitem [{\citenamefont {Santos}\ and\ \citenamefont
  {Rigol}(2010{\natexlab{a}})}]{santos_rigol_10b}%
  \BibitemOpen
  \bibfield  {author} {\bibinfo {author} {\bibfnamefont {L.~F.}\ \bibnamefont
  {Santos}}\ and\ \bibinfo {author} {\bibfnamefont {M.}~\bibnamefont {Rigol}},\
  }\bibfield  {title} {\enquote {\bibinfo {title} {Localization and the effects
  of symmetries in the thermalization properties of one-dimensional quantum
  systems},}\ }\href {\doibase 10.1103/PhysRevE.82.031130} {\bibfield
  {journal} {\bibinfo  {journal} {Phys. Rev. E}\ }\textbf {\bibinfo {volume}
  {82}},\ \bibinfo {pages} {031130} (\bibinfo {year}
  {2010}{\natexlab{a}})}\BibitemShut {NoStop}%
\bibitem [{\citenamefont {Neuenhahn}\ and\ \citenamefont
  {Marquardt}(2012)}]{neuenhahn_marquardt_12}%
  \BibitemOpen
  \bibfield  {author} {\bibinfo {author} {\bibfnamefont {C.}~\bibnamefont
  {Neuenhahn}}\ and\ \bibinfo {author} {\bibfnamefont {F.}~\bibnamefont
  {Marquardt}},\ }\bibfield  {title} {\enquote {\bibinfo {title}
  {Thermalization of interacting fermions and delocalization in {F}ock
  space},}\ }\href {\doibase 10.1103/PhysRevE.85.060101} {\bibfield  {journal}
  {\bibinfo  {journal} {Phys. Rev. E}\ }\textbf {\bibinfo {volume} {85}},\
  \bibinfo {pages} {060101} (\bibinfo {year} {2012})}\BibitemShut {NoStop}%
\bibitem [{\citenamefont {Khatami}\ \emph {et~al.}(2013)\citenamefont
  {Khatami}, \citenamefont {Pupillo}, \citenamefont {Srednicki},\ and\
  \citenamefont {Rigol}}]{khatami_pupillo_13}%
  \BibitemOpen
  \bibfield  {author} {\bibinfo {author} {\bibfnamefont {E.}~\bibnamefont
  {Khatami}}, \bibinfo {author} {\bibfnamefont {G.}~\bibnamefont {Pupillo}},
  \bibinfo {author} {\bibfnamefont {M.}~\bibnamefont {Srednicki}}, \ and\
  \bibinfo {author} {\bibfnamefont {M.}~\bibnamefont {Rigol}},\ }\bibfield
  {title} {\enquote {\bibinfo {title} {Fluctuation-dissipation theorem in an
  isolated system of quantum dipolar bosons after a quench},}\ }\href {\doibase
  10.1103/PhysRevLett.111.050403} {\bibfield  {journal} {\bibinfo  {journal}
  {Phys. Rev. Lett.}\ }\textbf {\bibinfo {volume} {111}},\ \bibinfo {pages}
  {050403} (\bibinfo {year} {2013})}\BibitemShut {NoStop}%
\bibitem [{\citenamefont {Steinigeweg}\ \emph {et~al.}(2013)\citenamefont
  {Steinigeweg}, \citenamefont {Herbrych},\ and\ \citenamefont
  {Prelov\ifmmode~\check{s}\else \v{s}\fi{}ek}}]{steinigeweg_herbrych_13}%
  \BibitemOpen
  \bibfield  {author} {\bibinfo {author} {\bibfnamefont {R.}~\bibnamefont
  {Steinigeweg}}, \bibinfo {author} {\bibfnamefont {J.}~\bibnamefont
  {Herbrych}}, \ and\ \bibinfo {author} {\bibfnamefont {P.}~\bibnamefont
  {Prelov\ifmmode~\check{s}\else \v{s}\fi{}ek}},\ }\bibfield  {title} {\enquote
  {\bibinfo {title} {Eigenstate thermalization within isolated spin-chain
  systems},}\ }\href {\doibase 10.1103/PhysRevE.87.012118} {\bibfield
  {journal} {\bibinfo  {journal} {Phys. Rev. E}\ }\textbf {\bibinfo {volume}
  {87}},\ \bibinfo {pages} {012118} (\bibinfo {year} {2013})}\BibitemShut
  {NoStop}%
\bibitem [{\citenamefont {Beugeling}\ \emph {et~al.}(2014)\citenamefont
  {Beugeling}, \citenamefont {Moessner},\ and\ \citenamefont
  {Haque}}]{beugeling_moessner_14}%
  \BibitemOpen
  \bibfield  {author} {\bibinfo {author} {\bibfnamefont {W.}~\bibnamefont
  {Beugeling}}, \bibinfo {author} {\bibfnamefont {R.}~\bibnamefont {Moessner}},
  \ and\ \bibinfo {author} {\bibfnamefont {Masudul}\ \bibnamefont {Haque}},\
  }\bibfield  {title} {\enquote {\bibinfo {title} {Finite-size scaling of
  eigenstate thermalization},}\ }\href {\doibase 10.1103/PhysRevE.89.042112}
  {\bibfield  {journal} {\bibinfo  {journal} {Phys. Rev. E}\ }\textbf {\bibinfo
  {volume} {89}},\ \bibinfo {pages} {042112} (\bibinfo {year}
  {2014})}\BibitemShut {NoStop}%
\bibitem [{\citenamefont {Kim}\ \emph {et~al.}(2014)\citenamefont {Kim},
  \citenamefont {Ikeda},\ and\ \citenamefont {Huse}}]{kim_14}%
  \BibitemOpen
  \bibfield  {author} {\bibinfo {author} {\bibfnamefont {H.}~\bibnamefont
  {Kim}}, \bibinfo {author} {\bibfnamefont {T.~N.}\ \bibnamefont {Ikeda}}, \
  and\ \bibinfo {author} {\bibfnamefont {D.~A.}\ \bibnamefont {Huse}},\
  }\bibfield  {title} {\enquote {\bibinfo {title} {Testing whether all
  eigenstates obey the eigenstate thermalization hypothesis},}\ }\href
  {\doibase 10.1103/PhysRevE.90.052105} {\bibfield  {journal} {\bibinfo
  {journal} {Phys. Rev. E}\ }\textbf {\bibinfo {volume} {90}},\ \bibinfo
  {pages} {052105} (\bibinfo {year} {2014})}\BibitemShut {NoStop}%
\bibitem [{\citenamefont {Sorg}\ \emph {et~al.}(2014)\citenamefont {Sorg},
  \citenamefont {Vidmar}, \citenamefont {Pollet},\ and\ \citenamefont
  {Heidrich-Meisner}}]{sorg_vidmar_14}%
  \BibitemOpen
  \bibfield  {author} {\bibinfo {author} {\bibfnamefont {S.}~\bibnamefont
  {Sorg}}, \bibinfo {author} {\bibfnamefont {L.}~\bibnamefont {Vidmar}},
  \bibinfo {author} {\bibfnamefont {L.}~\bibnamefont {Pollet}}, \ and\ \bibinfo
  {author} {\bibfnamefont {F.}~\bibnamefont {Heidrich-Meisner}},\ }\bibfield
  {title} {\enquote {\bibinfo {title} {Relaxation and thermalization in the
  one-dimensional {B}ose-{H}ubbard model: A case study for the interaction
  quantum quench from the atomic limit},}\ }\href {\doibase
  10.1103/PhysRevA.90.033606} {\bibfield  {journal} {\bibinfo  {journal} {Phys.
  Rev. A}\ }\textbf {\bibinfo {volume} {90}},\ \bibinfo {pages} {033606}
  (\bibinfo {year} {2014})}\BibitemShut {NoStop}%
\bibitem [{\citenamefont {Rigol}(2014)}]{rigol_14}%
  \BibitemOpen
  \bibfield  {author} {\bibinfo {author} {\bibfnamefont {M.}~\bibnamefont
  {Rigol}},\ }\bibfield  {title} {\enquote {\bibinfo {title} {Quantum quenches
  in the thermodynamic limit},}\ }\href {\doibase
  10.1103/PhysRevLett.112.170601} {\bibfield  {journal} {\bibinfo  {journal}
  {Phys. Rev. Lett.}\ }\textbf {\bibinfo {volume} {112}},\ \bibinfo {pages}
  {170601} (\bibinfo {year} {2014})}\BibitemShut {NoStop}%
\bibitem [{\citenamefont {Tang}\ \emph {et~al.}(2015)\citenamefont {Tang},
  \citenamefont {Iyer},\ and\ \citenamefont {Rigol}}]{tang_iyer_15}%
  \BibitemOpen
  \bibfield  {author} {\bibinfo {author} {\bibfnamefont {B.}~\bibnamefont
  {Tang}}, \bibinfo {author} {\bibfnamefont {D.}~\bibnamefont {Iyer}}, \ and\
  \bibinfo {author} {\bibfnamefont {M.}~\bibnamefont {Rigol}},\ }\bibfield
  {title} {\enquote {\bibinfo {title} {Quantum quenches and many-body
  localization in the thermodynamic limit},}\ }\href {\doibase
  10.1103/PhysRevB.91.161109} {\bibfield  {journal} {\bibinfo  {journal} {Phys.
  Rev. B}\ }\textbf {\bibinfo {volume} {91}},\ \bibinfo {pages} {161109(R)}
  (\bibinfo {year} {2015})}\BibitemShut {NoStop}%
\bibitem [{sup()}]{supmat}%
  \BibitemOpen
  \href@noop {} {}\bibinfo {note} {See Supplemental Material for more
  information about the weak symmetry breaking fields, quantum chaos
  indicators, and observables.}\BibitemShut {Stop}%
\bibitem [{\citenamefont {Aubry}\ and\ \citenamefont
  {Andr{\'e}}(1980)}]{Aubry1980}%
  \BibitemOpen
  \bibfield  {author} {\bibinfo {author} {\bibfnamefont {S.}~\bibnamefont
  {Aubry}}\ and\ \bibinfo {author} {\bibfnamefont {G.}~\bibnamefont
  {Andr{\'e}}},\ }\bibfield  {title} {\enquote {\bibinfo {title} {Analyticity
  breaking and anderson localization in incommensurate lattices},}\ }\href@noop
  {} {\bibfield  {journal} {\bibinfo  {journal} {Ann. Israel Phys. Soc}\
  }\textbf {\bibinfo {volume} {3}},\ \bibinfo {pages} {18} (\bibinfo {year}
  {1980})}\BibitemShut {NoStop}%
\bibitem [{\citenamefont {Atas}\ \emph {et~al.}(2013)\citenamefont {Atas},
  \citenamefont {Bogomolny}, \citenamefont {Giraud},\ and\ \citenamefont
  {Roux}}]{Atas2013}%
  \BibitemOpen
  \bibfield  {author} {\bibinfo {author} {\bibfnamefont {Y.~Y.}\ \bibnamefont
  {Atas}}, \bibinfo {author} {\bibfnamefont {E.}~\bibnamefont {Bogomolny}},
  \bibinfo {author} {\bibfnamefont {O.}~\bibnamefont {Giraud}}, \ and\ \bibinfo
  {author} {\bibfnamefont {G.}~\bibnamefont {Roux}},\ }\bibfield  {title}
  {\enquote {\bibinfo {title} {Distribution of the ratio of consecutive level
  spacings in random matrix ensembles},}\ }\href@noop {} {\bibfield  {journal}
  {\bibinfo  {journal} {Phys. Rev. Lett.}\ }\textbf {\bibinfo {volume} {110}},\
  \bibinfo {pages} {084101} (\bibinfo {year} {2013})}\BibitemShut {NoStop}%
\bibitem [{\citenamefont {Lev}\ \emph {et~al.}(2015)\citenamefont {Lev},
  \citenamefont {Cohen},\ and\ \citenamefont {Reichman}}]{Lev2015}%
  \BibitemOpen
  \bibfield  {author} {\bibinfo {author} {\bibfnamefont {Y.~Bar}\ \bibnamefont
  {Lev}}, \bibinfo {author} {\bibfnamefont {G.}~\bibnamefont {Cohen}}, \ and\
  \bibinfo {author} {\bibfnamefont {David~R.}\ \bibnamefont {Reichman}},\
  }\bibfield  {title} {\enquote {\bibinfo {title} {Absence of diffusion in an
  interacting system of spinless fermions on a one-dimensional disordered
  lattice},}\ }\href@noop {} {\bibfield  {journal} {\bibinfo  {journal} {Phys.
  Rev. Lett.}\ }\textbf {\bibinfo {volume} {114}},\ \bibinfo {pages} {100601}
  (\bibinfo {year} {2015})}\BibitemShut {NoStop}%
\bibitem [{\citenamefont {Santos}\ and\ \citenamefont
  {Rigol}(2010{\natexlab{b}})}]{santos_rigol_10a}%
  \BibitemOpen
  \bibfield  {author} {\bibinfo {author} {\bibfnamefont {L.~F.}\ \bibnamefont
  {Santos}}\ and\ \bibinfo {author} {\bibfnamefont {M.}~\bibnamefont {Rigol}},\
  }\bibfield  {title} {\enquote {\bibinfo {title} {Onset of quantum chaos in
  one-dimensional bosonic and fermionic systems and its relation to
  thermalization},}\ }\href {\doibase 10.1103/PhysRevE.81.036206} {\bibfield
  {journal} {\bibinfo  {journal} {Phys. Rev. E}\ }\textbf {\bibinfo {volume}
  {81}},\ \bibinfo {pages} {036206} (\bibinfo {year}
  {2010}{\natexlab{b}})}\BibitemShut {NoStop}%
\bibitem [{\citenamefont {Bauer}\ \emph {et~al.}(2015)\citenamefont {Bauer},
  \citenamefont {Dorfner},\ and\ \citenamefont
  {Heidrich-Meisner}}]{bauer_dorfner_15}%
  \BibitemOpen
  \bibfield  {author} {\bibinfo {author} {\bibfnamefont {A.}~\bibnamefont
  {Bauer}}, \bibinfo {author} {\bibfnamefont {F.}~\bibnamefont {Dorfner}}, \
  and\ \bibinfo {author} {\bibfnamefont {F.}~\bibnamefont {Heidrich-Meisner}},\
  }\bibfield  {title} {\enquote {\bibinfo {title} {Temporal decay of {N}\'eel
  order in the one-dimensional {F}ermi-{H}ubbard model},}\ }\href {\doibase
  10.1103/PhysRevA.91.053628} {\bibfield  {journal} {\bibinfo  {journal} {Phys.
  Rev. A}\ }\textbf {\bibinfo {volume} {91}},\ \bibinfo {pages} {053628}
  (\bibinfo {year} {2015})}\BibitemShut {NoStop}%
\bibitem [{\citenamefont {Gramsch}\ and\ \citenamefont
  {Rigol}(2012)}]{gramsch_rigol_12}%
  \BibitemOpen
  \bibfield  {author} {\bibinfo {author} {\bibfnamefont {C.}~\bibnamefont
  {Gramsch}}\ and\ \bibinfo {author} {\bibfnamefont {M.}~\bibnamefont
  {Rigol}},\ }\bibfield  {title} {\enquote {\bibinfo {title} {Quenches in a
  quasidisordered integrable lattice system: Dynamics and statistical
  description of observables after relaxation},}\ }\href {\doibase
  10.1103/PhysRevA.86.053615} {\bibfield  {journal} {\bibinfo  {journal} {Phys.
  Rev. A}\ }\textbf {\bibinfo {volume} {86}},\ \bibinfo {pages} {053615}
  (\bibinfo {year} {2012})}\BibitemShut {NoStop}%
\bibitem [{\citenamefont {Iyer}\ \emph {et~al.}(2015)\citenamefont {Iyer},
  \citenamefont {Srednicki},\ and\ \citenamefont {Rigol}}]{iyer_srednicki_15}%
  \BibitemOpen
  \bibfield  {author} {\bibinfo {author} {\bibfnamefont {D.}~\bibnamefont
  {Iyer}}, \bibinfo {author} {\bibfnamefont {M.}~\bibnamefont {Srednicki}}, \
  and\ \bibinfo {author} {\bibfnamefont {M.}~\bibnamefont {Rigol}},\ }\bibfield
   {title} {\enquote {\bibinfo {title} {Optimization of finite-size errors in
  finite-temperature calculations of unordered phases},}\ }\href {\doibase
  10.1103/PhysRevE.91.062142} {\bibfield  {journal} {\bibinfo  {journal} {Phys.
  Rev. E}\ }\textbf {\bibinfo {volume} {91}},\ \bibinfo {pages} {062142}
  (\bibinfo {year} {2015})}\BibitemShut {NoStop}%
\end{thebibliography}

\end{document}